\documentclass[aps,pra,reprint,groupedaddress]{revtex4-1}

\usepackage{amsmath,amssymb,graphicx}
\usepackage{array}
\usepackage[caption=false]{subfig}
\usepackage{bm}
\usepackage{amsfonts}
\usepackage{epsfig,epstopdf}
\usepackage{subfig}
\usepackage{url}

\def\tK{\tilde K}

\def\Tr{{\rm Tr}}

\def\la{\left\langle}
\def\ra{\right\rangle}

\def\bq{{\bf q}}

\def\bu{{\bf u}}

\def\hrho{\hat \rho}

\def\bH{{\bf H}}

\def\be{\begin{equation}}
\def\ee{\end{equation}}
\def\ba{\begin{align}}
\def\nn{\nonumber\\}

\def\defeq{\buildrel \rm def \over =}

\def\bs{{\bf s}}
\def\b0{{\bf 0}}

\def\la{\left\langle}
\def\ra{\right\rangle}

\def\bs{{\bf s}}
\def\bsperp{\bs_\perp}
\def\partiall{\partial^{(l)}}
\def\partials{\partial^{(s)}}

\def\bu{{\bf u}}

\def\bH{{\bf H}}

\def\bl{\bm{l}}
\def\blperp{{\bm{l}}_\perp}

\def\Tr{{\rm Tr\ }}

\def\be{\begin{equation}}
\def\ee{\end{equation}}
\def\ba{\begin{align}}
\def\nn{\nonumber\\}
\def\la{\langle}
\def\ra{\rangle}

\def\defeq{\buildrel \rm def \over =}

\def\Re{{\rm Re}}
\def\Im{{\rm Im}}

\begin{document}

\title{Quantum limited super-localization and super-resolution of a source pair in three dimensions}

\author{Sudhakar Prasad}
\altaffiliation{Also at School of Physics and Astronomy, University of Minnesota, Minneapolis, MN 55455}

\email{sprasad@unm.edu}
\author{Zhixian Yu}
\affiliation{Department of Physics
and Astronomy, University of New Mexico, Albuquerque, New Mexico 87131}
\date{\today}

\pacs{(100.6640) Superresolution; (110.3055) Information theoretical analysis;
(110.6880) Three-dimensional image acquisition; (110.7348) Wavefront encoding; (110.1758)
Computational imaging; (270.5585) Quantum information and processing}

\begin{abstract}

A recent paper \cite{YuPrasad18} considered the problem of quantum limited estimation of the separation vector for a pair of incoherent point sources in all three dimensions.  Here we extend our analysis to treat the problem of simultaneously estimating the location of the centroid and the separation of the source pair, which is equivalent to localizing both sources simultaneously. We first calculate the quantum Fisher information for simultaneous pair centroid-separation estimation and then discuss the fundamental, estimation-theoretic trade-offs between the two tasks, which we confirm using simulations.  
\end{abstract}

\vspace{-1cm}


\maketitle

Optical superresolution imaging has been a subject of great current interest, ranging from single-molecule localization imaging using uncorrelated photons from randomly photoactivated, well separated individual molecules \cite{Rust06} to quantum-correlated, optical centroid measuring states \cite{Tsang09, Schwartz13, Unternahrer18}  to the use of wavefront projections \cite{Tsang16, Nair16, Paur16,Tham17,Chrostowski17,Zhou18,Tsang18}.   

A recent paper \cite{YuPrasad18} by the present authors has extended the analysis of quantum limited estimation of the separation of a pair of incoherent point sources from one \cite{Tsang16,Nair16} and two \cite{Ang17} transverse dimensions to include the third, axial dimension in the photon-counting limit. The quantum limit on the variance of unbiased estimation of the three-dimensional (3D) separation vector, as determined by the inverse of the quantum Fisher information (QFI) \cite{Helstrom76,Braunstein94,Paris09}, may be expressed most simply, as we showed, in terms of the correlation of the wavefront phase gradients in the imaging aperture. Because of the linearity of the wavefront phase with respect to (w.r.t) the pair-separation vector, QFI and its inverse, the quantum Cram\'er-Rao bound (QCRB), are both independent of that vector.

In the present Communication, we extend our work further to calculate QFI and QCRB for the joint estimation of the position of the centroid and the separation of a pair of equally bright sources in the photon-counting (Poisson) limit in all three spatial dimensions. This analysis is more general than that of Ref.~\cite{Napoli18} in which the authors restrict the localization of the two sources jointly to a single transverse dimension and the line-of-sight dimension. Furthermore, our analysis, like our previous paper's \cite{YuPrasad18}, makes no assumptions about the aperture geometry, such as inversion symmetry that other papers on quantum-limited pair superresolution problem have used to derive their results.

The QFI matrix, $\bH$, is defined to have elements $H_{\mu\nu}\defeq \Re \Tr(\hrho \hat L_\mu \hat L_\nu)$, where $\Re$ denotes the real part and $\hat L_\mu$ is the symmetric logarithmic derivative (SLD), w.r.t. the $\mu$th parameter, of the density operator $\hrho$,
\be
\label{rho}
\hrho ={1\over 2}\left(|\tilde K_+\ra\la \tK_+|+|\tK_-\ra\la \tK_-|\right),
\ee
for a photon emitted by the incoherent source pair and captured by the imaging aperture. The six parameters, $l_x,l_y,l_z$ and $s_x,s_y,s_z$ of interest here are the three Cartesian components of the normalized pair-separation and pair-centroid position vectors, $\bl$ and $\bs$, respectively, with $\bs$ defined in the same way as $\bl$ is in Ref.~\cite{YuPrasad18}. The two pure single-photon states, $|\tK_\pm\ra$, are emitted by the two point sources located at $\bs\pm \bl$, respectively. The corresponding normalized wavefunctions have the following representations over the aperture \cite{Goodman17}:
\ba
\label{wavefunction}
\la \bu| \tK_\pm\ra = &\exp(\pm i\phi_0)\, P(\bu)\, \exp(-i2\pi\bsperp\cdot\bu-i\pi s_z u^2)\nn
&\times \exp[\mp i\Psi(\bu;\bl)],
\end{align}
in which $P(\bu)$ is a generally-complex pupil function obeying the normalization condition,
\be
\label{normalize}
\int d^2u\, |P(\bu)|^2 = 1,
\ee 
the phase function, $\Psi(\bu;\bl)$, has the form,
\be
\label{pairphase}
\Psi(\bu;\bl)=2\pi \bu\cdot\blperp +\pi u^2l_z,   
\ee
and the phase constant, $\phi_0$,
is conveniently chosen to make the inner product, $\Delta\defeq \la \tK_-|\tK_+\ra$, real. In view of relations (\ref{wavefunction}) and (\ref{pairphase}) for the wavefunction and $\Psi$, this inner product may be expressed as
\be
\label{Delta}
\Delta =\exp(-2i\phi_0)\int d^2u\, |P({\bf u})|^2\exp(i4\pi{\bf l}_\perp\cdot{\bf u}+i2\pi l_z u^2),
\ee
which like the phase constant, $\phi_0$, is independent of the centroid position vector, ${\bf s}$. 
For the clear, unit-radius circular aperture, $P(\bu)$ is simply $1/\sqrt{\pi}$ times the indicator function for the aperture. Due to form ({\ref{wavefunction}) of the wavefunctions, $\Delta$ does not depend on $\bs$.

For the problem of estimating $\bl$ alone, QFI matrix elements were shown in Ref.~\cite{YuPrasad18} to have the form,
\be
\label{QFIll}
H^{(ll)}_{\mu\nu}=4\left[\langle \partiall_\mu\Psi\partiall_\nu\Psi\rangle-\langle \partiall_\mu\Psi\rangle\langle\partiall_\nu\Psi\rangle\right],
\ee 
where angular brackets here denote weighted aperture averages, with $|P(\bu)|^2$ being the weight function. 


The minimum error of joint estimation of $\bl$ and $\bs$ is given by the inverse of a $6\times 6$ QFI matrix of which $\bH^{(ll)}$ given by expression (\ref{QFIll}) may be regarded as a $3\times 3$ diagonal block. The full QFI matrix may be organized as a collection of four $3\times 3$ blocks,
\be
\label{QFIblockform}
\bH=\left(
\begin{array}{c|c}
\bH^{(ll)}&\bH^{(ls)}\\
\hline
\bH^{(sl)} &\bH^{(ss)}
\end{array}
\right),
\ee
with matrix elements defined by the formula
\ba
\label{QFIblockelements}
H^{(ab)}_{\mu\nu}&=H^{(ba)}_{\nu\mu}\nn
                             &=\Re \Tr(\hrho \hat L_\mu^{(a)}\hat L^{(b)}_\nu);\ a,b=l,s; \ \mu,\nu=x,y,z.
\end{align}  

The remaining matrix elements, $H_{\mu\nu}^{(ls)}, H_{\mu\nu}^{(ss)}$, follow from their general form \cite{YuPrasad18},
\begin{align}
\label{QFIblock}
H_{\mu\nu}^{(ab)}&=\sum_{i=\pm}{1\over e_i}\partial_\mu^{(a)} e_i\partial_\nu^{(b)} e_i\nn
&+4\Re \sum_{i=\pm}{1\over e_i}(\partial_\mu^{(a)} \langle 
e_i|)(\hat\rho-e_i\hat I)^2\partial_\nu^{(b)}|e_i\rangle\nn
&+4\Delta^2\Re \sum_{i\neq j}\left({1\over e_i}-e_i\right)\langle e_i|\partial_\mu^{(a)} |e_j\rangle\langle e_j|\partial_\nu^{(b)} |e_i\rangle,
\end{align} 
in which $\partial_\mu^{(l)}\defeq\partial/\partial l_\mu$ and $\partial_\mu^{(s)}\defeq\partial/\partial s_\mu$ denote partial derivatives relative to $l_\mu$ and $s_\mu$, respectively, and $\hat I$ is the identity operator. The eigenvalues, $e_\pm$, and associated orthonormal eigenstates, $|e_\pm\ra$, are easily derived,
\be
\label{eigen}
e_\pm = {1\pm \Delta\over 2}, \ |e_\pm\ra ={1\over \sqrt{2(1\pm \Delta)}}\left(|\tK_+\ra\pm |\tK_-\ra\right).
\ee
Since $\hrho=e_+|e_+\ra\la e_+|+e_-|e_-\ra\la e_-|$, we may write
\ba
\label{rel1}
(\hrho-e_+\hat I)\partial_\nu |e_+\ra =&e_+[|e_+\ra\la e_+|\partial_\nu |e_+\ra -\partial_\nu|e_+\ra]\nn
                                                            &+e_-|e_-\ra\la e_-|\partial_\nu|e_+\ra, 
\end{align}
in which $\partial_\nu$ denotes a partial derivative w.r.t. any of the six parameters being estimated.
Multiplying Eq.~(\ref{rel1}) by its Hermitian adjoint (h.a.) on the left, with $\nu$ replaced by $\mu$ in the latter, we reach one of the two inner products occurring in the middle sum of expression (\ref{QFIblock}). Two of the nine terms of which this product is comprised vanish from the orthogonality of the eigenstates, $\la e_+|e_-\ra=0$. Two other terms cancel out identically, and the remaining five combine neatly into a set of three distinct terms,
\ba
\label{QFIblockmiddleterm1}
(\partial_\mu \langle e_+|)&(\hat\rho-e_+\hat I)^2\partial_\nu|e_+\rangle=-(e_-^2-2e_+e_-)\la e_+|\partial_\mu|e_-\ra\nn
&\times\la e_-|\partial_\nu|e_+\ra
+e_+^2\la e_+|\partial_\mu|e_+\ra\la e_+|\partial_\nu|e_+\ra\nn
&+e_+^2(\partial_\mu\la e_+|) \partial_\nu|e_+\ra.
\end{align}
Noting that $\hrho$ is formally invariant under an interchange of the $+$ and $-$ subscripts in relation (\ref{QFIblockmiddleterm1}) yields the second inner product in the second sum,
\ba
\label{QFIblockmiddleterm2}
(\partial_\mu \langle e_-|)&(\hat\rho-e_-\hat I)^2\partial_\nu|e_-\rangle=-(e_+^2-2e_+e_-)\la e_-|\partial_\mu|e_+\ra\nn
&\times\la e_+|\partial_\nu|e_-\ra
+e_-^2\la e_-|\partial_\mu|e_-\ra\la e_-|\partial_\nu|e_-\ra\nn
&+e_-^2(\partial_\mu\la e_-|) \partial_\nu|e_-\ra.
\end{align}  

Since $\Delta$ does not depend on $\bs$, taking the partial derivative of $|e_+\ra$, given by expression (\ref{eigen}), w.r.t. any component of $\bs$, and taking the inner product of the resulting expression with the bra $\la e_\pm|$, obtained by taking the h.a. of expression (\ref{eigen}), generates the following useful identities:
\ba
\label{identity2}
\la e_+|\partials_\mu|e_+\ra &={\la \tK_+|\partials_\mu|\tK_+\ra+i\Im \la \tK_+|\partials_\mu|\tK_-\ra\over (1+\Delta)};\nn
\la e_-|\partials_\mu|e_+\ra &={\Re \la \tK_+|\partials_\mu|\tK_-\ra\over \sqrt{1-\Delta^2}}.
\end{align}
To arrive at these identities, we used the relations, $\la \tK_+|\partials_\mu|\tK_+\ra=\la \tK_-|\partials_\mu|\tK_-\ra$ and $\la \tK_+|\partials_\mu|\tK_-\ra=-\la \tK_-|\partials_\mu|\tK_+\ra^*$, that follow from form (\ref{wavefunction}) of the states $|\tK_\pm\ra$ and from the fact that $\partials_\mu(\la\tK_-|\tK_+\ra)=0$, respectively. 
The identities,
\be
\label{identity1}
\la e_+|\partial_\mu^{(l)}|e_+\ra =0,\ \la e_-|\partial_\mu^{(l)}|e_+\ra = {1\over\sqrt{1-\Delta^2}}\la \tK_+|\partial_\mu |\tK_+\ra,
\ee
proved similarly in the supplemental notes of Ref.~\cite{YuPrasad18}, and four more obtained by the interchange of $|e_+\ra$ and $|e_-\ra$ in Eqs.~(\ref{identity2}) and (\ref{identity1}), which entails the substitutions $|\tK_\pm\ra\to \pm |\tK_\pm\ra$ and $\Delta\to -\Delta$ according to expressions (\ref{eigen}) for $|e_\pm\ra$, namely
\ba
\label{identity3}
\la e_-|\partials_\mu|e_-\ra &={\la \tK_+|\partials_\mu|\tK_+\ra-i\Im \la \tK_+|\partials_\mu|\tK_-\ra\over(1-\Delta)},\nn
\la e_+|\partials_\mu|e_-\ra &=-{\Re \la \tK_+|\partials_\mu|\tK_-\ra\over \sqrt{1-\Delta^2}},
\end{align}
and
\be
\label{identity4}
\la e_-|\partial_\mu^{(l)}|e_-\ra =0,\ \la e_+|\partial_\mu^{(l)}|e_-\ra = {1\over\sqrt{1-\Delta^2}}\la \tK_+|\partial_\mu |\tK_+\ra,
\ee
comprise the full set of identities that can simplify expression (\ref{QFIblock}) for the elements of the blocks $\bH^{(sl)}$ and $\bH^{(ss)}$.

Since $e_\pm$ are independent of $\bs$, it follows that the first sum on the right hand side in expression (\ref{QFIblock}) vanishes identically, while the other two sums may be combined into one in view of expressions (\ref{QFIblockmiddleterm1}) and (\ref{QFIblockmiddleterm2}) for the two terms of the second sum. Using the identities, $e_\mp^2-2e_+e_-=\Delta^2-e_\pm^2$, we may thus obtain the following expression for the block $\bH^{(sl)}$:
\begin{align}
\label{QFIblocksl1}
H_{\mu\nu}^{(sl)}&=4(1-\Delta^2)\Re \sum_{i\neq j}e_i\langle e_i|\partials_\mu |e_j\rangle\la e_j|\partiall_\nu|e_i\ra\nn
&+4\Re \sum_{i=\pm}e_i(\partial_\mu^{(s)} \la 
e_i|)\partiall_\nu|e_i\ra.
\end{align} 
From identities (\ref{identity1})-(\ref{identity4}), we see that $\la e_\pm|\partials_\mu|e_\mp\ra$ are real, while $\la e_\pm|\partiall_\nu|e_\mp\ra$ are purely imaginary, the latter since $\la |\tK_+|\partiall_\mu|\tK_+\ra$ is purely imaginary on account of the form (\ref{wavefunction}) of the wavefunctions. Consequently, the first term in expression (\ref{QFIblocksl1}) vanishes identically. That the second sum there - and thus the entire off-diagonal QFI block, $\bH^{(sl)}$ - also vanishes,
\be
\label{QFIblocksl3}
\bH^{(sl)}=0,
\ee
is shown in \cite{Supp2}.
In other words, there is no increase of the minimum error of unbiased joint estimation of the pair centroid-location and separation vectors over that of unbiased independent estimation of the two vectors. 

We turn now to $\bH^{(ss)}$, which entails some of the same calculational steps as $\bH^{(sl)}$.
The main difference, however, is that $\la e_-|\partials_\nu|e_+\ra$ is purely real, unlike the purely imaginary $\la e_-|\partiall_\nu|e_+\ra$, so the analog of the first term in expression (\ref{QFIblocksl1}) for $H^{(sl)}_{\mu\nu}$ no longer vanishes for $H^{(ss)}_{\mu\nu}$. After some algebra \cite{Supp2}, we reach the following expression for $\bH^{(ss)}$:
\ba 
\label{QFIblockss1}
H_{\mu\nu}^{(ss)}&=4\Big[(\partials_\mu\la \tK_+|) \partials_\nu|\tK_+\ra\nn
&-\Re\la \tK_+|\partials_\mu|\tK_-\ra\Re\la \tK_+|\partials_\nu|\tK_-\ra\Big]\nn
&-{4\over 1-\Delta^2}\Big(\Im\la\tK_+|\partial_\mu^{(s)}|\tK_+ \ra \Im\la\tK_+|\partial_\nu^{(s)}|\tK_+ \ra \nn
&+\Im\la \tK_+|\partials_\mu|\tK_-\ra\Im \la \tK_+|\partials_\nu|\tK_-\ra\Big)\nn
&+{4\Delta\over 1-\Delta^2}\Big(\Im\la\tK_+|\partial_\mu^{(s)}|\tK_+ \ra \Im\la\tK_+|\partial_\nu^{(s)}|\tK_- \ra \nn
&+\Im\la \tK_+|\partials_\nu|\tK_+\ra\Im \la \tK_+|\partials_\mu|\tK_-\ra\Big).
\end{align} 
In Eq.~(\ref{QFIblockss1}), all matrix elements involving only $|\tK_+\ra$ and its derivatives, but not $|\tK_-\ra$, are easily evaluated as simple aperture averages of powers of aperture coordinates, while the matrix element $\la \tK_+|\partials_\mu|\tK_-\ra$ may be evaluated in the aperture plane using the wavefunctions (\ref{wavefunction}) and $\Delta$ given by relation (\ref{Delta}), 
\ba
\label{mixedelement}
\la \tK_+|\partials_\mu&|\tK_-\ra= -{\exp(-i2\phi_0)\over 2\pi}\int_A d^2 u\, \nn
                                                   &\times\partiall_\mu[\exp(4i\pi\bu\cdot\blperp+2i\pi u^2 l_z)]\nn
=&-{\Delta\int_A d^2 u\,\partiall_\mu[\exp(4i\pi\bu\cdot\blperp+2i\pi u^2 l_z)]\over
                              2\int_A d^2 u\,  \exp(4i\pi\bu\cdot\blperp+2i\pi u^2 l_z)}.
\end{align}                              

Expression (\ref{QFIblockss1}) for QFI for estimating the centroid location coordinates alone is independent of those coordinates. This is fundamentally a consequence of the global translational invariance of a shift-invariant imager, as the centroid location vector, $\bs$,  can be changed by an arbitrary additive constant vector by a mere change of the origin of the coordinate system, under which the pair separation vector, $\bl$, is invariant. Physically speaking, an axial refocusing and a transverse alignment of the imager are all that are needed to place the pair centroid at the origin in the source space, an action that cannot affect the fidelity with which the centroid can be estimated. This QFI depends only on $\bl$ through $\Delta$ and certain aperture integrals. 

The off-diagonal elements of ${\bf H}^{(ss)}$ do not vanish, which reflects the interdependence of the errors of estimation of the three coordinates of the centroid location when estimating them jointly. This is in sharp contrast to the three components of the pair-separation vector, which can be estimated independently of each other \cite{YuPrasad18}.

Since the overall QFI matrix (\ref{QFIblockform}) is block diagonal, its inverse is obtained by inverting each diagonal block,
\be
\label{QCRBblockform}
\bH^{-1}=\left(
\begin{array}{c|c}
\big(\bH^{(ll)}\big)^{-1}&{\bf 0}\\
\hline
{\bf 0} &\big(\bH^{(ss)}\big)^{-1}
\end{array}
\right),
\ee
in which $\big(\bH^{(ll)}\big)^{-1}$ has the value \cite{YuPrasad18}, 
\be
\label{QCRBll}
\big(\bH^{(ll)})^{-1}=\left(
\begin{array}{ccc}
{1\over 4\pi^2}&0&0\\
0&{1\over 4\pi^2}&0\\
0&0&{3\over \pi^2}
\end{array}
\right).
\ee

Specializing to the case of the imaging aperture being clear and circular, we numerically evaluated the elements (\ref{QFIblockss1}) of the QFI matrix $\bH^{(ss)}$ and then inverted it to compute the values of QCRB for estimating the centroid location coordinates.  In Fig.~\ref{QCRBsx_vs_lx}, we plot QCRB for estimating $s_x$ vs $l_x$ 
for a number of different values of the other transverse component of the pair-separation vector, namely $l_y$. The curves start out close to the source-localization QCRB of $1/(4\pi^2)\approx 0.0253$ when the two sources are close to each other and thus approximate a single source. They also asymptote toward the same QCRB value for large separations, since in this limit sources can be localized individually and their centroid thus determined to the same precision as their individual positions. For intermediate values of $l_x$, the minimum error variance for estimating $s_x$ is increased due to the image blur caused by a finite aperture size when the sources are transversely not well separated on the Abbe-Rayleigh scale, $l_\perp\lesssim 0.25$. Changing $l_z$, the axial separation of the pair, from a small value of 0.025 to 0.25 does not improve the $s_x$ estimation error significantly, as seen in the small difference between the curves in the left and right panels. Because of perfect $x\leftrightarrow y$ symmetry for a circular aperture, an identical behavior was confirmed by our numerical evaluation of QCRB for the estimation of $s_y$ vs. $l_y$.

\begin{figure}[htb]
\centerline{\includegraphics[width=0.95\columnwidth]{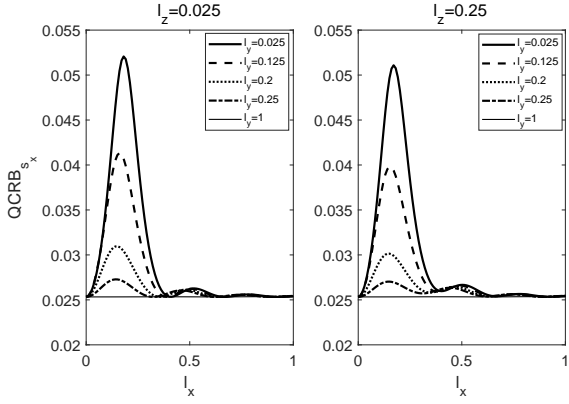}}
\vspace{-0.2cm}
\caption{Plots of QCRB for $s_x$ vs. $l_x$ for two different values of $l_z$, namely 0.025 (left panel) and 0.25 (right panel)}
\label{QCRBsx_vs_lx}
\end{figure}

In Fig.~\ref{QCRBsx_vs_ly}, we display QCRB for estimating $s_x$ vs. $l_y$.
As expected, with increasing $l_y$, the minimum error variance for estimating $s_x$ decreases as the sources get farther apart in the orthogonal direction. Once again, as the sources get well separated, when either $l_x$ or $l_y$ or both become large, the minimum error variance for locating the pair centroid in the transverse plane approaches the localization QCRB, namely 0.0253. The relative vertical positions of the curves for different values of $l_x$ are consistent with the peaks seen in Fig.~\ref{QCRBsx_vs_lx}.

\begin{figure}[htb]
\centerline{\includegraphics[width=0.95\columnwidth]{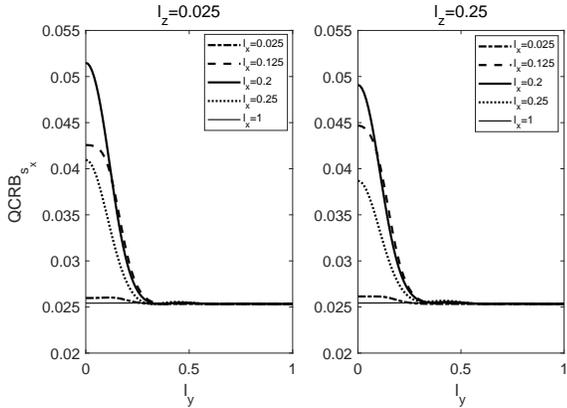}}
\vspace{-0.2cm}
\caption{Plots of QCRB for $s_x$ vs. $l_y$ for two different values of $l_z$, namely 0.025 (left panel) and 0.25 (right panel)}
\label{QCRBsx_vs_ly}
\end{figure}

In Fig.~\ref{QCRBsz}, we plot QCRB for estimating $s_z$, the axial coordinate of the pair centroid, as a function of $l_z$, the axial component of the pair-separation vector. The intrinsic imprecision of estimating the axial coordinate, as reflected in the larger axial-localization QCRB of $3/\pi^2\approx 0.304$ than the transverse-localization QCRB of 0.0253, is seen in the larger scatter, at the two ends of small and large axial separations, among plots for different values of $l_\perp$, the transverse separation. Interestingly, there are multiple values of $l_z$ for which QCRB for estimating $s_z$ has minima at the localization QCRB of 0.304 with increasing $l_z$. The larger QCRB for $s_z$ than that for $s_x$ or $s_y$ has to do with the quadratic, rather than linear, dependence of the aperture phase on axial coordinates, which implies a lower overall first-order differential sensitivity of wavefront projections to them. This fact also accounts for why the horizontal scale of the plots for axial-coordinate estimation is larger than that for transverse-coordinate estimation plotted in previous figures.
\begin{figure}[htb]
\centerline{\includegraphics[width=0.95\columnwidth]{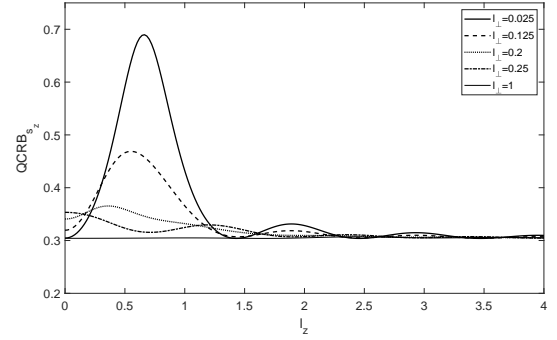}}
\vspace{-0.2cm}
\caption{Plots of QCRB for $s_z$ vs. $l_z$ for several values of $l_\perp$}
\label{QCRBsz}
\end{figure}

For small pair separations, the pair centroid can be localized by standard image based methods to a precision comparable to QCRB, but coherent projections are necessary to attain quantum limited estimation of the pair separation. We envisage a two-arm experimental approach, similar to that of Ref.~\cite{Tsang16}, in which a beam splitter directs, on average, a preset fraction of photons into one arm in which a 3D localization imager like a rotating-PSF imager \cite{Pavani08,Lew11,Prasad13,YuPrasad16}, an astigmatic imager \cite{Huang08}, a multiplane imager \cite{Ram08}, or a radial shearing interferometer \cite{Backlund18} is placed. The remaining photons traverse a second arm that has the same holographic aperture-plane filter as that described in Ref.~\cite{YuPrasad18}, namely $\sum_n Z_n(\bu)\cos\bq_n\cdot\bu$, in which $Z_n$ denotes the $n$th Zernike polynomial \cite{Noll76} and $\bq_n$ is the transverse offset wavevector of the $n$th mode.

We show results of a partial simulation of this approach to estimate the pair separation using the maximum-likelihood (ML) estimator described in \cite{YuPrasad18}, subject to a certain centroid localization error achieved in the centroid localization arm and a fixed number, $M$, of photons in the holographic filter arm. The photons divide into the various pure-Zernike channels according to the probabilities, $\{P_n\defeq \la Z_n|\hrho|Z_n\ra\mid n=1,\ldots,N\}$, and into the unmeasured channels with probability, $\bar P=1-\sum_{n=1}^N P_n$, to yield a multinomial distribution of observed counts from which the ML estimator can extract the separation vector. The classical FI matrix elements \cite{VT68,Kay93} for estimating the three pair-separation coordinates from the multinomial distribution of counts take the per-photon form \cite{Supp2},
\be
\label{FI_multi}
J_{\mu\nu}^{(ll)}/M = \sum_{n=1}^N {(\partiall_\mu P_n)\,(\partiall_\nu P_n)\over P_n}+{(\partiall_\mu \bar P)\,(\partiall_\nu\bar P)\over \bar P},
\ee
which was evaluated by numerical integration for $N=4$. 

In Fig.~4 (a), we plot the variance of the ML estimation of $l_x$ obtained from a sample of 40 draws of $\bs$ from a product-Gaussian statistical distribution with zero means and standard deviations, $\sigma^{(s)}_x=\sigma^{(s)}_y=0.005, \ \sigma^{(s)}_z=0.01$, with 400 multinomial data frames for each such $\bs$ sample and with $10^6$ photons per frame. The mean and standard deviation of these estimation variances over the 40 $\bs$ draws are denoted by the square symbols and error bars through them. The classical CRB, which is the $xx$ diagonal element of the inverse of the FI matrix (\ref{FI_multi}), when averaged over the 40 $\bs$ draws, is shown by the dot-dash curve and that for $\bs=0$ by the solid curve in the figure. The results of simulation track well this last curve, presumably since for simulated data we take $\bs=0$ when extracting the estimates of $\bl$. The divergence of the dot-dash curve for $l_x\to 0$ is due to the fact that for $s_x\neq 0$, neither $Z_2$ nor another pure Zernike is an exclusively matched filter \cite{Turin60} for $l_x$ in the limit $l_x\to 0$. For most of the range of $l_x$ away from 0, however, the four Zernike projections furnish excellent convergence of the variance of the separation estimate based on them to QCRB. Because of the azimuthal symmetry of the optical system and our choice of the Zernikes, the same results as shown in this figure also hold for the estimation of $l_y$. 

In Fig.~4 (b), we display analogous curves for estimating the axial separation, $l_z$. An important difference from the estimation of lateral separation is that all classical CRB curves diverge in the limit $l_z\to 0$, as no Zernike provides an exclusively matched filter for the azimuthally symmetric defocus phase, as we noted in Ref.~\cite{YuPrasad18}.  All CRB curves asymptote toward the QCRB line, however, as $l_z$ grows in value.   
\begin{figure}[htb]
\centering
\subfloat[]
{\includegraphics[width=0.49\columnwidth, height=0.55\columnwidth]{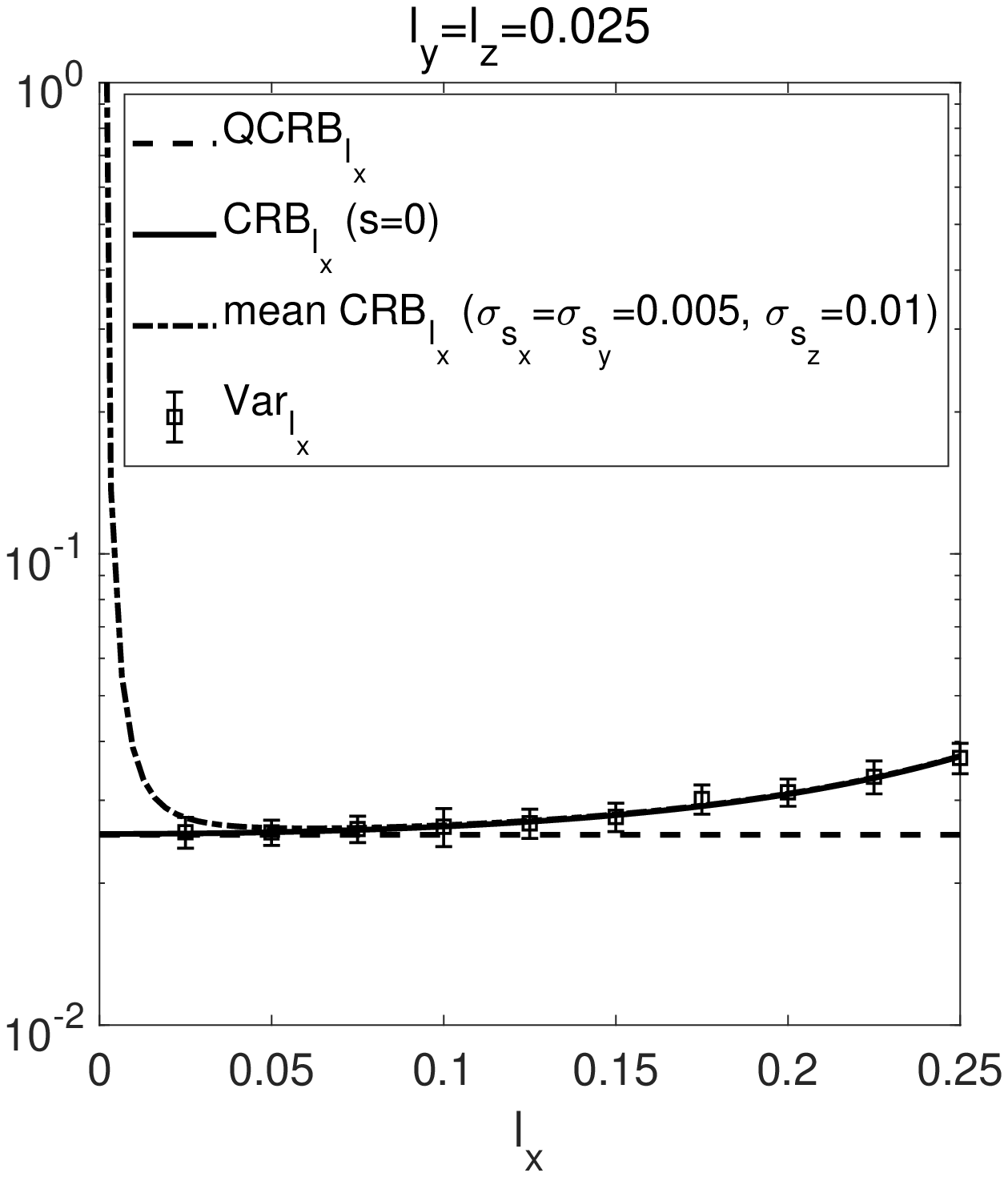}}\
\subfloat[]
{\includegraphics[width=0.49\columnwidth, height=0.55 \columnwidth]{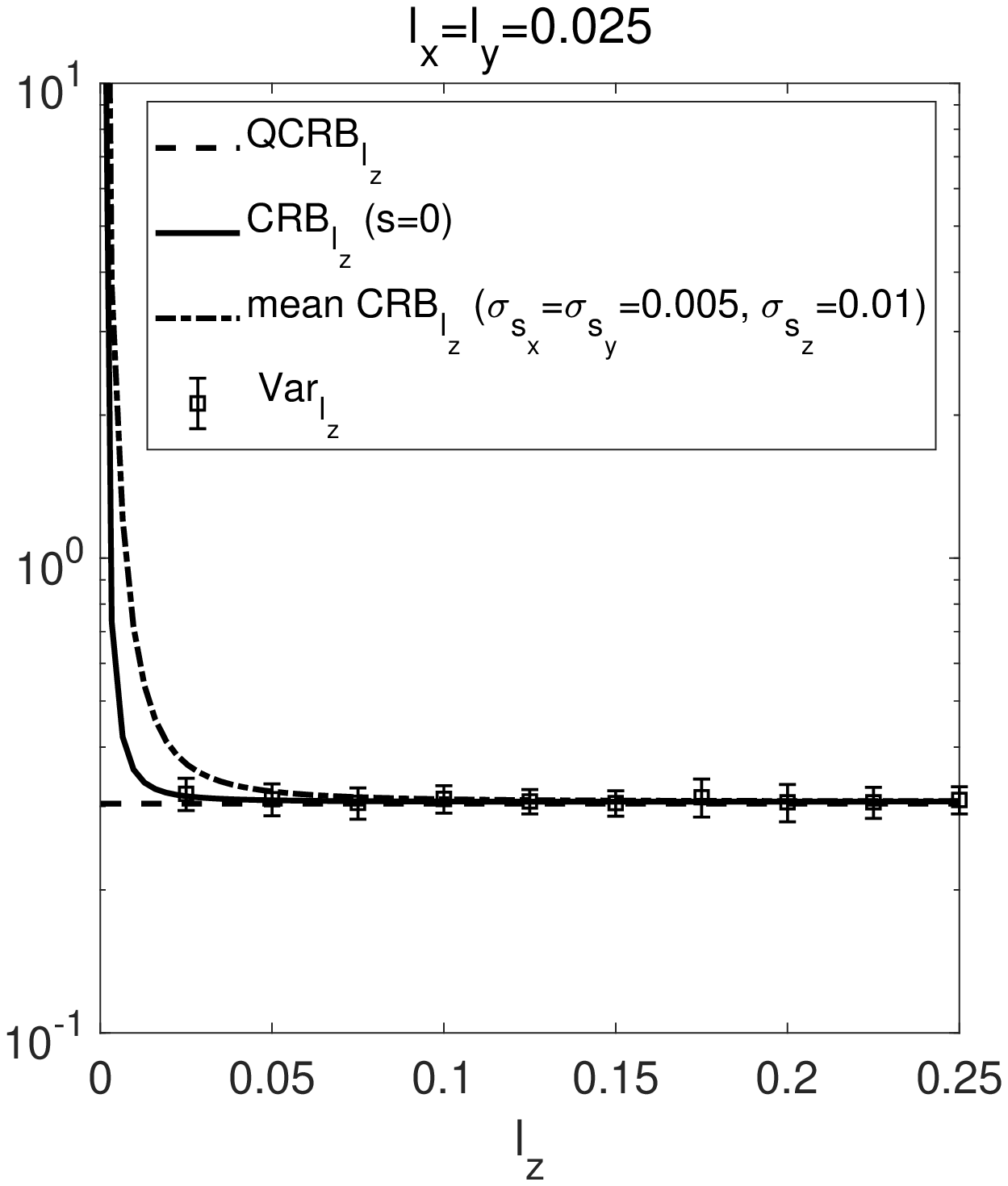}}
\caption{\label{fig:f1} (a) Plot of variance of estimation of $l_x$ with changing values of $l_x$, with the other two $l$ coordinates being equal to 0.025, for $\sigma_{s_x}=\sigma_{s_y}=0.005;\ \sigma_{s_z}=0.01$; (b) Same as (a) except $l_x\to l_z$.}
\end{figure}   

This Communication has extended our previous analysis of quantum limited source pair separation to include 3D localization of the pair centroid as well. While no fundamental bounds on estimator variances can depend on the centroid coordinates for a spatially invariant system like the one we have considered, any uncertainties in their estimation, for which image-based methods suffice, affect the estimation variances of the pair separation. 

\section*{Acknowledgments}
The work was partially supported by the US Air Force Office of Scientific Research under grant no. FA9550-15-1-0286.

\end{document}


\title{Supplemental Material}

\author{Sudhakar Prasad and Zhixian Yu}
\date{\today}
\maketitle








\section{Vanishing of the off-diagonal QFI Block, ${\bf H}^{(sl)}$}

In Eq.~(18), main text, the first term of the right-hand side (RHS) vanishes, as we already noted. We may simplify the second term there by noting that $\partials_\mu|e_\pm\rangle$ and $\partiall_\nu|e_\pm\rangle$, when the form (9), main text, of the eigenstates is used, may be written as
\ba
\label{partial_der1}
\partials_\mu \langle e_\pm|=& {1\over \sqrt{2(1\pm \Delta)}}\left(\partials_\mu \langle\tK_+| \pm  \partials_\mu \langle\tK_-|\right);\nn
\partiall_\nu |e_\pm\rangle=& \mp{\partiall_\nu \Delta\over 2(1\pm\Delta)}|e_\pm\rangle +{1\over \sqrt{2(1\pm \Delta)}}\left(\partiall_\nu |\tK_+\rangle \pm  \partiall_\nu |\tK_-\rangle\right),
\end{align}
in which we used the fact that $\Delta$ is independent of the centroid vector, $\bs$, to arrive at the first line.
Taking the inner product of the above two states, multiplying the product by $e_\pm=(1\pm\Delta)/2$, and then adding the two terms that result corresponding to the upper and lower signs, we may express the second sum in Eq.~(18), main text, as
\be
\label{second_term}
\sum_{i=\pm}e_i(\partial_\mu^{(s)} \la e_i|)\partiall_\nu|e_i\ra =  {1\over 2}\left[(\partials_\mu\la \tK_+|) \partiall_\nu|\tK_+\ra+ (\partials_\mu\la \tK_-|) \partiall_\nu|\tK_-\ra\right]-{\partiall_\nu \Delta\over 4}\left[(\partials_\mu\langle e_+|)|e_+\ra-(\partials_\mu\langle e_-|)|e_-\ra\right],
\ee
where the terms inside the second bracket follow from the expression for $\partials_\mu\la e_\pm|$ given in Eq.~(\ref{partial_der1}). From the form of the wavefunctions (2), main text, it follows that the two terms inside the first bracket on the RHS of Eq.~(\ref{second_term}) are exactly negative of each other, so their sum vanishes, which simplifies Eq.~(\ref{second_term}) to the form  
\be
\label{second_term1}
\sum_{i=\pm}e_i(\partial_\mu^{(s)} \la e_i|)\partiall_\nu|e_i\ra =  -{\partiall_\nu \Delta\over 4}\left[(\partials_\mu\langle e_+|)|e_+\ra-(\partials_\mu\langle e_-|)|e_-\ra\right].
\ee 
Since the eigenstates are normalized, $\la e_\pm |e_\pm\ra=1$, we have the identity, $\partials_\mu(\la e_\pm|e_\pm\ra) = 0$, which from the product rule of differentiation is equivalent to the relation, 
\be
\label{par_der_ident}
(\partials_\mu \la e_\pm|)|e_\pm\ra=-\la e_\pm |\partials_\mu|e_\pm\ra.
\ee
Using the complex-conjugation property of the inner product, we may write the left-hand side of Eq.~(\ref{par_der_ident}) as $\la e_\pm|\partials_\mu|e_\pm\ra^*$, which when equated to its RHS implies that $(\partials_\mu\la e_\pm|)|e_\pm\ra$ is purely imaginary. Consequently, expression (\ref{second_term1}) is purely imaginary, and thus $H^{(sl)}_{\mu\nu}$, which is the proportional to its real part, vanishes identically,
\be
\label{allODblock}
H^{(sl)}_{\mu\nu}=0.
\ee

\section{Derivation of centroid-localization QFI}\label{sec:QFIderivation}

The matrix elements of the centroid-localization QFI, $\bH^{(ss)}$, are given by replacing all $\partiall$ derivatives by $\partials$  in Eq.~(16) of the main paper and then adding the sum, $\sum_{i=\pm}e_i \la e_i|\partials_\mu|e_i\ra\la e_i|\partials_\nu|e_i\ra$, arising from the non-vanishing second terms on the RHS of Eqs.~(10) and (11) of the main paper, 
\be
\label{QFIblockss1}
H_{\mu\nu}^{(ss)}=4(1-\Delta^2)\Re \sum_{i\neq j}e_i\langle e_i|\partials_\mu |e_j\rangle\la e_j|\partials_\nu|e_i\ra+4\Re \sum_{i=\pm}e_i\big[\la e_i|\partials_\mu|e_i\ra\la e_i|\partials_\nu|e_i\ra+(\partial_\mu^{(s)} \la 
e_i|)\partials_\nu|e_i\ra\big].
\ee 
The matrix elements, $\la e_+|\partials_\mu|e_\pm\ra$ and $\la e_-|\partials_\mu|e_\pm\ra$, were already evaluated in  the main paper in terms of those involving the pure emission states, $|\tK_\pm\ra$, as
\ba
\label{identity2}
\la e_+|\partials_\mu|e_+\ra &={\la \tK_+|\partials_\mu|\tK_+\ra+i\Im \la \tK_+|\partials_\mu|\tK_-\ra\over (1+\Delta)};\ \la e_-|\partials_\mu|e_-\ra ={\la \tK_+|\partials_\mu|\tK_+\ra-i\Im \la \tK_+|\partials_\mu|\tK_-\ra\over (1-\Delta)}\nn
\la e_-|\partials_\mu|e_+\ra &={\Re \la \tK_+|\partials_\mu|\tK_-\ra\over \sqrt{1-\Delta^2}}=-\la e_+|\partials_\mu|e_-\ra.
\end{align}
The remaining matrix elements, $(\partials_\mu\la e_\pm|)\partials_\nu|e_\pm\ra$, are obtained by taking appropriate derivatives of the following expressions for $|e_\pm\ra$ in terms of the pure emission states:
\be
\label{eigen}
|e_\pm\ra = {1\over \sqrt{2(1\pm \Delta)}}\big(|\tK_+\ra\pm |\tK_-\ra\big), 
\ee
and noting that $\Delta$ is independent of all centroid-location coordinates. These matrix elements may thus be expressed as
 \be
 \label{identity3}
 (\partials_\mu\la e_\pm|)\partials_\nu|e_\pm\ra={1\over 2(1\pm \Delta)}\left(\partials_\mu \la \tK_+|\pm\partials_\mu \la \tK_-|\right)\left(\partials_\nu|\tK_+\ra\pm\partials_\nu|\tK_-\ra\right).
 \ee
 
Since $e_\pm=(1/2)(1\pm \Delta)$, substituting the last of the matrix elements in Eq.~(\ref{identity2}) into the first sum in Eq.~(\ref{QFIblockss1}) reduces it to the form,
\be
\label{firstsum}
4(1-\Delta^2)\Re \sum_{i\neq j}e_i\langle e_i|\partials_\mu |e_j\rangle\la e_j|\partials_\nu|e_i\ra=-4\Re \la \tK_+|\partials_\mu|\tK_-\ra\Re \la \tK_+|\partials_\nu|\tK_-\ra.
\ee
Substituting the first two of the matrix elements in Eq.~(\ref{identity3}) into the first part of the second sum on the RHS of Eq.~(\ref{QFIblockss1}) and then taking its real part evaluates it to the form,
\ba
\label{secondsum1}
4\Re \sum_{i=\pm}e_i\la e_i|\partials_\mu|e_i\ra&\la e_i|\partials_\nu|e_i\ra\nn
=&-{2\over 1+\Delta}\left(\Im\la \tK_+|\partials_\mu|\tK_+\ra +\Im \la \tK_+|\partials_\mu|\tK_-\ra\right)\left(\Im\la \tK_+|\partials_\nu|\tK_+\ra +\Im \la \tK_+|\partials_\nu|\tK_-\ra\right)\nn
&-{2\over 1-\Delta}\left(\Im\la \tK_+|\partials_\mu|\tK_+\ra- \Im \la \tK_+|\partials_\mu|\tK_-\ra\right)\left(\Im\la \tK_+|\partials_\nu|\tK_+\ra- \Im \la \tK_+|\partials_\nu|\tK_-\ra\right)\nn
=&-{4\over 1-\Delta^2}\left(\Im\la \tK_+|\partials_\mu|\tK_+\ra\ \Im \la \tK_+|\partials_\nu|\tK_+\ra
                                        +\Im\la \tK_+|\partials_\mu|\tK_-\ra\ \Im \la \tK_+|\partials_\nu|\tK_-\ra\right)\nn
 &+{4\Delta \over 1-\Delta^2}\left(\Im\la \tK_+|\partials_\mu|\tK_+\ra\ \Im \la \tK_+|\partials_\nu|\tK_-\ra
                                        +\Im\la \tK_+|\partials_\nu|\tK_+\ra\ \Im \la \tK_+|\partials_\mu|\tK_-\ra\right),
\end{align}
in which we used the fact that $\la\tK_\pm|\partials_\mu|\tK_\pm\ra$ are purely imaginary quantities.
Finally, substituting the matrix element (\ref{identity3}) into the second part of the second sum in Eq.~(\ref{QFIblockss1}) also simplifies it,
 \ba
 \label{secondsum2}
 4\Re \sum_{i=\pm}e_i(\partial_\mu^{(s)} \la e_i|)\partials_\nu|e_i\ra = &2\big[(\partial_\mu^{(s)} \la \tK_+|)\partials_\nu|\tK_+\ra+(\partial_\mu^{(s)} \la \tK_-|)\partials_\nu|\tK_-\ra\big]\nn
 =& 4(\partial_\mu^{(s)} \la \tK_+|)\partials_\nu|\tK_+\ra,
 \end{align}
 in which we used the fact that the matrix elements, $(\partial_\mu^{(s)} \la \tK_\pm|)\partials_\nu|\tK_\pm\ra$, are both real and equal to each other as both wavefunctions $\la\bu|\tK_\pm\ra$ are pure exponential phase functions over the aperture, with an identical dependence on the centroid location vector, $\bs$. 
Substituting expressions (\ref{firstsum})-(\ref{secondsum2}) into Eq.~(\ref{QFIblockss1}) generates the final expression for the centroid-localization QFI, $\bH^{(ss)}$, that we use in the main paper. 

\section{Derivation of FI for Multinomial Photocount Distribution}
For $(N+1)$ projection channels, with per-photon probabilities being $P_1,\ldots,P_{N+1}$, in which $P_{N+1}\defeq \bar P = 1-\sum_{n=1}^N P_n$, the probability, $P(m_1,\ldots,m_{N+1})$, of detecting $m_1,\ldots,m_{N+1}$ photons in those channels when a total of $M$ photons are incident on the projection system is given by the multi-nomial distribution (MND),
\be
\label{MN_PD}
P(m_1,\ldots,m_{N+1})=M!\prod_{n=1}^{N+1} {P_n^{m_n}\over m_n!}\Theta(m_1,\ldots,m_{N+1}),
\ee
with $\Theta$ denoting the indicator function for the discrete space of constraints defined as
\be
\label{MN_constraints}
\sum_{n=1}^{N+1} m_n=M,\ \ m_1,\ldots, m_n=0,1,\ldots, M.
\ee
The channel probabilities, $P_1,\ldots,P_N$, depend on the six parameters being estimated in the present problem.

Taking the logarithm of expression (\ref{MN_PD}) and the partial derivatives of the resulting expression with respect to the $\mu$th and $\nu$th parameters successively, then multiplying the resulting expressions with each other, and finally taking the expectation of their product over MND yields the following form for the $\mu\nu$ matrix element of the associated FI:
\ba
\label{MN_FI1}
J_{\mu\nu} &= \sum_{n=1}^{N+1}\sum_{l=1}^{N+1}\la m_n m_l\ra{(\partial_\mu\ln P_n)\, (\partial_\nu\ln P_l)}\nn
                  &=M(M-1)\sum_{n=1}^{N+1} \sum_{l=1}^{N+1}P_n P_l{(\partial_\mu\ln P_n)\, (\partial_\nu\ln P_l)}+M\sum_{n=1}^{N+1}P_n         
                  (\partial_\mu\ln P_n)\, (\partial_\nu\ln P_n)\nn
                  &=M(M-1) \left[\sum_{n=1}^{N+1}P_n (\partial_\mu\ln P_n)\right]\left[\sum_{l=1}^{N+1}P_l (\partial_\nu\ln P_l)\right]\nn
                  &\quad +M\sum_{n=1}^{N+1}P_n (\partial_\mu\ln P_n)\, (\partial_\nu\ln P_n)\nn
                  &=M\sum_{n=1}^{N+1}{ (\partial_\mu P_n)\, (\partial_\nu P_n)\over P_n},
\end{align}
in which we used the well known formula for the second moment of MND,
\be
\label{MN_moment}
\la m_n m_l\ra =M(M-1) P_n P_l +MP_n\delta_{nl},
\ee
to reach the second line and the fact that since $\sum_{n=1}^{N+1} P_n =1$, any partial derivative of it vanishes,
\be
\label{MN_ident}
\sum_{n=1}^{N+1} {P_n(\partial_\mu \ln P_n)}=0,
\ee
to arrive at the final expression.